# Bayesian-Optimized Multi-Source Domain Adaptation for Post-Earthquake Damage Assessment

Y. Zhang[1] and X. Liang[2]

## ABSTRACT
Efficient and intelligent post-earthquake structural damage assessment is critical for rapid disaster response. Although data-driven approaches have shown promise in this domain, traditional supervised learning relies on large labeled datasets that are impractical to obtain for earthquake-damaged structures. To overcome this limitation, we propose a Bayesian-optimized multi-source domain adaptation framework for predicting post-earthquake structural damage on a target building without the need for any damage labels. The framework comprises three key steps. First, it extracts features from multiple source domains and the target domain and feeds them into a classifier and a domain discriminator. The classifier ensures the features remain damage-sensitive, while the discriminator promotes their invariance across domains. Second, the framework assigns a weighing factor to each source domain to balance their contributions during training. Finally, Bayesian optimization is employed to optimize these source domain weights, aiming to maximize prediction accuracy on the target domain. This framework offers a robust solution for structural damage assessment when labeled data are scarce, significantly enhancing post-earthquake damage assessment capabilities.

## Introduction
Timely and reliable quantification of earthquake-induced structural damage is fundamental for protecting human life, guiding emergency operations, and formulating effective recovery strategies [1]. Within hours of a major seismic event, authorities must identify buildings at risk of collapse, deploy rescue teams, and allocate temporary shelter and repair resources in an informed manner. For decades, these decisions have relied primarily on manual visual inspection [2,3]: professional engineers conduct systematic, building-by-building surveys, recording cracks, residual drifts, and other signs of distress before assigning safety tags. Although well established, this practice is inherently slow and labor-intensive, and it exposes inspectors to considerable danger, especially since aftershocks can trigger further structural degradation. Subjective judgements also lead to variability between inspectors, and the process is difficult to scale to large metropolitan areas with thousands of buildings requiring rapid appraisal.

Model-based numerical simulation provides a complementary approach to damage assessment. With sufficiently detailed as-built documentation, a nonlinear finite-element (FE) model updated with recorded ground motions can estimate inter-story drifts, plastic hinge rotations, and component stresses, thereby supporting objective and spatially resolved damage assessments [4]. In practice, however, reliable FE simulations depend on accurate material properties, boundary conditions, and input ground motions. This information is often incomplete or uncertain in the chaotic post-earthquake environment. Furthermore, thousands of nonlinear analyses must be executed to cover the range of possible excitations and parameter variations, imposing a computational burden that limits real-time applicability. The unavoidable modelling uncertainties also complicate the interpretation of numerical predictions in urgent decision contexts.

[1] Graduate Student Researcher, Zachry Dept. of Civil and Environmental Engineering, Texas A&M University, College Station, TX 77840 (email: zyf@tamu.edu)
[2] Assistant Professor, Zachry Dept. of Civil and Environmental Engineering, Texas A&M University, College Station, TX 77840 (email: xliang@tamu.edu)

These limitations have spurred a third line of research: data-driven structural-health monitoring (SHM). Rather than relying exclusively on physics-based models, data-driven approaches learn the mapping from measured dynamic response to damage state directly from historical observations. When labelled datasets exist, supervised machine-learning and deep learning models – such as support vector machines [5,6] and convolutional neural networks (CNNs) [7,8] – can achieve high classification accuracy. However, after a damaging earthquake, the very labels that supervised models require are unavailable, since generating those labels is the goal of the assessment. Moreover, a classifier trained on one structure often generalizes poorly to another because the distribution of features and damage labels varies with each building's geometry, materials, and boundary conditions. This domain transfer problem is particularly severe in civil engineering, where each building is effectively unique.

Domain adaptation (DA) offers a statistically principled solution by learning feature representations that remain sensitive to damage yet invariant to different domains. Early SHM applications of DA employed subspace-alignment techniques such as Transfer Component Analysis and Correlation Alignment, which project source and target data into a common latent space by minimizing kernelized statistical distances, typically the maximum mean discrepancy [9]. These methods improve robustness under moderate distribution differences. Nonetheless, they rely on manual feature extraction, cannot easily handle missing sensors or varying record lengths, and their performance deteriorates when the range of excitation intensities differs substantially between structures [10].

More recent advances introduce adversarial DA [11], wherein a feature extractor is trained to fool a domain discriminator while still maintaining high classification accuracy – a minimax optimization objective that encourages stronger distribution alignment [12,13]. Adversarial DA has improved performance in vibration-based damage assessment, three key challenges hinder its broader application for post-earthquake assessment. First, earthquake monitoring yields very long, multi-channel acceleration time series at high sampling frequency, making it non-trivial to extract compact and informative features from such data [14,15]. Second, post-earthquake assessments often require story-by-story damage estimates, not just a global building-level condition [16]. Third, in practice available labelled data may come from multiple source buildings that differ in height, stiffness, or construction quality. Balancing the influence of these disparate sources is essential to avoid negative transfer.

To address these challenges, we propose a Bayesian-optimized multi-source domain adaptation (BO-MSDA) framework that transfers structural damage knowledge without requiring target labels. Result shows that it achieves high story-level accuracy, offering a practical approach for rapid post-earthquake screening.

## Methodology

The proposed BO-MSDA framework transfers structural damage knowledge from labeled source buildings to an unlabeled target building. It integrates two main parts: domain adaptation network and Bayesian optimization. Through this integrated design, the BO-MSDA framework enables reliable, fully label-free story-level damage estimation. The overall framework is illustrated in Figure 1.

To overcome the difficulty of handling long, multi-channel acceleration records, we employ a feature extractor composed of a 1D-CNN and a Bi-LSTM encoder. The extracted representations are fed simultaneously to a damage classifier and multiple domain discriminators. The classifier ensures that the learned features remain damage-sensitive, while the discriminators align the feature distributions across domains to make them invariant to building differences. And feature extractors, discriminators, and classifiers together constitute domain adaptation networks. Notably, to prevent negative transfer when using multiple source buildings with different structural characteristics, we assign a weighting factor to each source domain in the domain adaptation network. These weights regulate each source's influence during training, allowing the framework to favor the most relevant sources while still exploiting the diversity of all the data.

Subsequently, because the choice of weights critically affects generalization performance, we employ

Bayesian optimization to automatically search the weight space. This process identifies the weight configuration that yields the highest prediction accuracy on the unlabeled target domain, ensuring that the transfer is both effective and reliable.

Finally, we evaluate the proposed methodology using a comprehensive simulation campaign comprising nonlinear time-history analyses of two reinforced-concrete moment frames (one 3-story and one 5-story) subjected to 5,400 recorded ground motions. All response–label pairs from the 3-story building serve as the labelled source domain, whereas each individual story of the 5-story building serves as a separate unlabeled target domain.

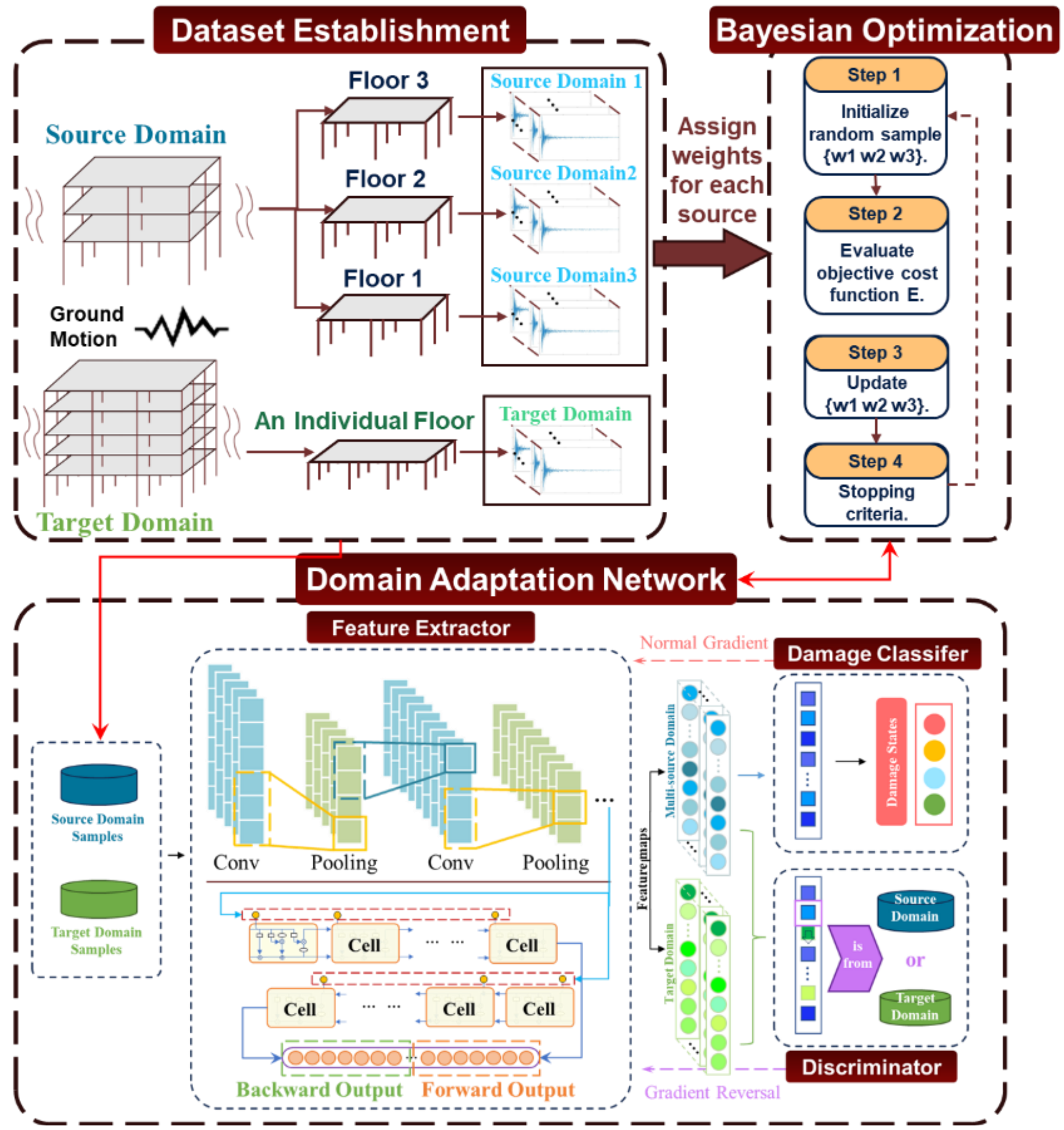

Figure 1. Framework of BO-MSDA.

## Case Study

### Dataset Establishment

The proposed damage assessment framework is demonstrated on two three-dimensional reinforced concrete special moment frames representative of modern U.S. practice: one 3-story structure and one 5-story structure. Both buildings have two bays in each principle direction, with 24 ft spans and a uniform story height of 14 ft. The concrete and reinforcing steel are modeled using the Concrete02 and Steel02 material models, respectively, and beams and columns are represented with fiber-based force elements. Floor slabs are treated as rigid diaphragms so that lateral response is governed by frame action. A magnitude-7 seismic scenario is considered, using 180 horizontal ground motion pairs from the PEER NGA-West2 database. Each ground motion is scaled to 30 intensity levels, producing a total of 5,400 motions per building. Nonlinear response history analyses are then conducted in parallel using OpenSees to efficiently perform these incremental dynamic analyses [17].

The simulations provide complete records of floor responses and inter-story deformations, forming the dataset for model training and evaluation. For each floor in the structure, we construct a feature vector from the acceleration time histories recorded at that floor and at the ceiling of that story (i.e., the floor above) in both the x and y directions. Acceleration measurements are chosen because they capture global structural behavior – including dynamic characteristics, energy dissipation, and vibration modes – and are readily obtainable in practice. The class label for each data sample is defined by inter-story drift ratios, a widely accepted indicator of structural damage: values below 1% correspond to class 0, between 1% and 2% to class 1, and above 2% to class 2 [18]. A larger drift ratio reflects greater accumulation of plastic deformation and therefore a more severe damage state.

**Results**

Overall, the BO-MSDA framework achieves high damage-state classification accuracy on the labeled 3-story source building and effectively transfers this knowledge to the 5-story target building without using any target labels. In the source domain, the confusion matrix exhibits strong diagonal dominance, confirming that the extracted features are highly damage-sensitive and yield reliable classification across all damage categories. When applied to the target domain, a natural drop in accuracy is observed due to structural differences in height, stiffness, and dynamic properties. Nevertheless, the model still preserves a consistent ranking of damage severity at each story level. Low-drift cases (Class 0) remain reliably recognized, and most moderate (Class 1) and severe (Class 2) cases are correctly identified. This outcome indicates that the learned features are effectively domain-invariant, even when applied to a different building.

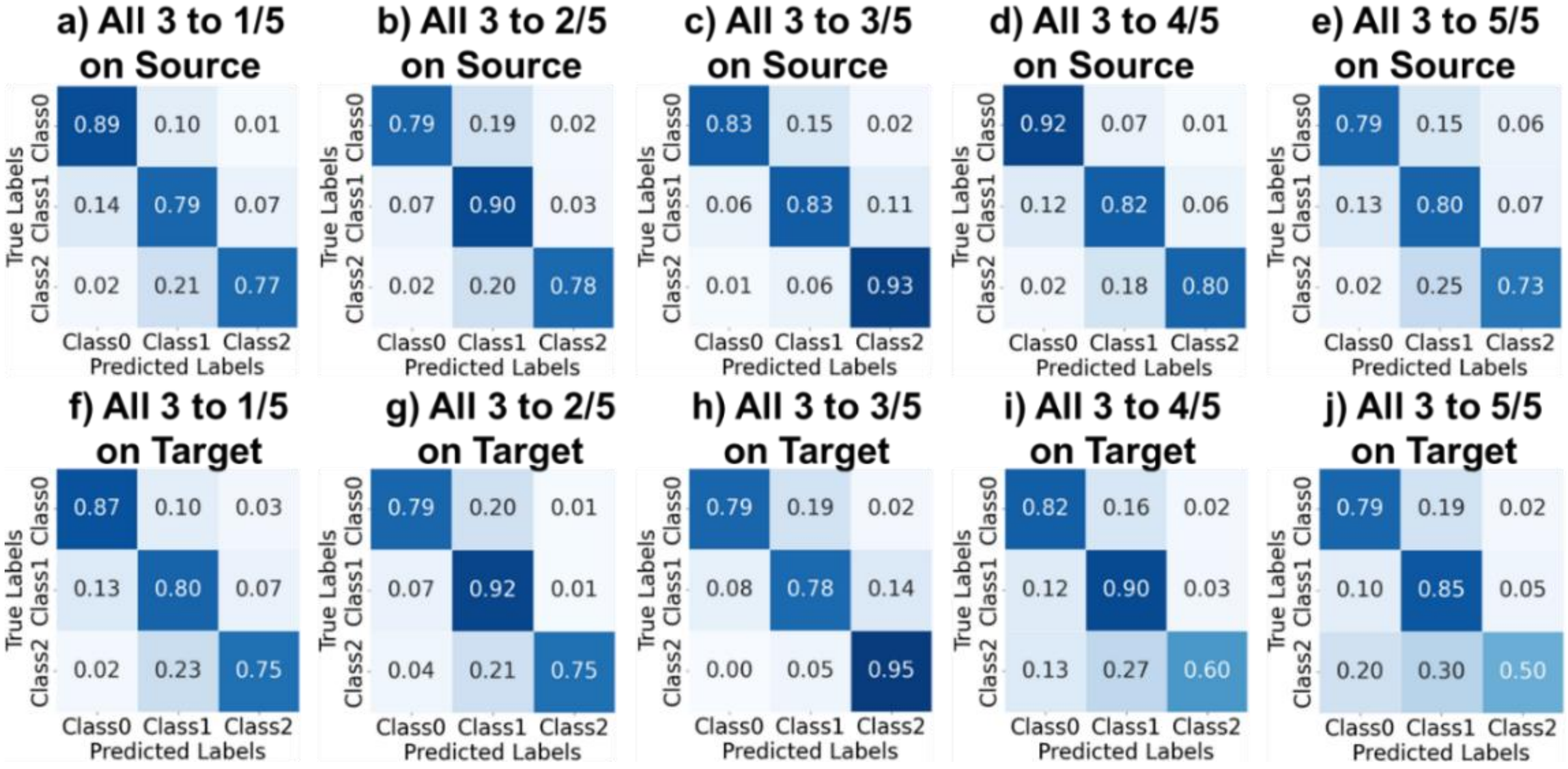

Figure 2. Confusion matrices for DA tasks, where "All 3 to 1/5" denotes all stories of the 3-story building as the source and the 1st story of the 5-story building as the target.

**Conclusions**

This work presented a BO-MSDA framework for story-level seismic damage assessment without requiring target labels. By combining adversarial domain adaptation network and Bayesian optimization of source weights, the method enables damage-sensitive knowledge transfer between buildings of different configurations. Numerical studies on 3 and 5-story reinforced concrete frames subjected to thousands of ground motion records confirmed that the framework not only maintains high accuracy on the labeled source but also generalizes effectively to the unlabeled target, preserving the relative ranking of damage severity across stories. These results highlight the potential of BO-MSDA as a practical tool for rapid, label-free post-earthquake screening, offering engineers timely guidance when detailed structural information is unavailable.